\documentclass[11pt]{article}

\usepackage{array}
\usepackage{epsfig}
\usepackage{graphics}

\title{\bf A mode elimination technique to improve convergence of iteration methods
 for finding solitary waves}

\author{ T.I. Lakoba\footnote{Corresponding author: lakobati@cems.uvm.edu, \ 1 (802) 656-2610} \ \ and
  J. Yang\footnote{jyang@cems.uvm.edu, \ 1 (802) 656-4314}
 \vspace{1cm} \\
  Department of Mathematics and Statistics, 16 Colchester Ave., \\
 University of Vermont, Burlington, VT 05401, USA}

\newcommand{\noi}{\noindent}
\newcommand{\D}{\Delta}
\newcommand{\be}{\begin{equation}}
\newcommand{\ee}{\end{equation}}
\newcommand{\ba}{\begin{array}}
\newcommand{\ea}{\end{array}}
\newcommand{\To}{\rightarrow}
\newcommand{\vecx}{{\bf x}}
\newcommand{\PM}{Petviashvili method}

\newcommand{\Int}{\int_{-\infty}^{\infty}}
\newcommand{\bea}{\begin{eqnarray}}
\newcommand{\eea}{\end{eqnarray}}

\newcommand{\tu}{\tilde{u}}

\begin{document}
\baselineskip 18 pt

\maketitle

\vskip 1 cm

We extend the key idea behind the generalized \PM\ of Ref. \cite{gP} 
by proposing a novel technique based on a similar idea. 
This technique systematically eliminates from the iteratively obtained 
solution a mode that is ``responsible" either for the divergence or the slow
convergence of the iterations. We demonstrate, theoretically and with examples,
that this mode elimination technique can be used both to 
obtain some nonfundamental solitary waves and to considerably accelerate convergence of
various iteration methods. As a collateral result, we compare the linearized iteration
operators for the generalized \PM\ and the well-known imaginary-time evolution method
and explain how their different structures account for the differences in the convergence
rates of these two methods.

\vskip 1.1 cm

\noi

\bigskip

\noi
{\bf Mathematical subject codes}: \ 35Qxx, 65B99, 65N99, 78A40, 78A99.

\newpage

\section{Introduction}

In the companion paper \cite{gP}, we proposed a generalization of the Petviashvili iteration
method for finding stationary solitary waves $u(\vecx)$ of scalar and vector Hamiltonian
equations with arbitrary form of nonlinearity:
\be
- M\,u + F(\vecx,u) =0, \qquad u(|\vecx|\To \infty) \To 0,
\label{p2_1_01}
\ee
where $M$ is a self-adjoint differential operator and, in the vector case, the nonlinear term
must satisfy a condition \ $ \partial F_i/\partial u_j = \partial F_j/\partial u_i $. 
(Recall that the original \PM\ \cite{Petviashvili76} was proposed for scalar equations with
power-law nonlinearity \ $F(\vecx,u)=u^p$.) A common form of operator $M$ (in the scalar case) is
\be
M=\mu -\nabla^2,
\label{p2_1_02}
\ee
where $\mu$ is the propagation constant of the solitary wave.
Thus, the generalized \PM, that obtains solutions with a specified propagation constant,
 can be applied to the same class of equations as the well-known
imaginary-time evolution method (ITEM) (see, e.g., \cite{GarciaRipollPG01}--\cite{YangL06})
that is used to find solitary waves with a specified power.

In the present work we extend the results of 
\cite{gP} as follows. In Section 2, we establish a mathematical relation between 
the generalized \PM\ and the ITEM. This discussion will also set the stage for the 
main result of this work, presented in Section 3. There, we develop the ideas behind the original
and generalized \PM s \cite{PelinovskyS04,gP} and propose a new technique that
we refer to as the mode elimination. This technique can be used to obtain nonfundamental (see below)
solitary waves, which the methods of \cite{gP}--\cite{YangL06} cannot obtain (the iterations would
diverge). However, since alternative methods of obtaining nonfundamental solitary waves exist
\cite{GarciaRipollPG01,SO},
we see the {\em main use} of the mode elimination in that it can considerably
accelerate convergence of various iteration methods. The corresponding examples are presented in 
Section 4, and the summary of our results is given in Section 5.



\setcounter{equation}{0}
\section{Convergence rates of the Petviashvili and the imaginary-time evolution methods}

In this Section, we will compare the convergence properties of the
generalized \PM\ \cite{gP} with those of the accelerated ITEM proposed in
Ref. \cite{YangL06}. This discussion will highlight a feature of the generalized
Petviashvili iteration scheme that will be important when we present our main
result --- the mode elimination technique --- in the next Section.
Of the two versions of the ITEM (with power and amplitude
normalizations) considered in \cite{YangL06},
we will focus on the one with power normalization, because its linearized operator can
be readily compared with that of the generalized \PM. 
In order not to obscure the main ideas by technical details, 
we restrict our presentation to the case of a single real-valued
equation (\ref{p2_1_01}) with $M$ given by Eq. (\ref{p2_1_02}), i.e., to:
\be
\nabla^2 u + F(\vecx,u)=\mu u\,.
\label{e5_01}
\ee

It is well-known that the convergence of an iteration method is determined by 
the properties of the {\em linearized} iteration equation. Namely, let
$u_n$ be the solution obtained at the $n$th iteration, and let the ``error" $\tu_n$ be defined as
\be
\tu_n=u_n-u, \quad |\tu_n| \ll |u|\,.
\label{p2_5_01}
\ee
As will be shown below, it satisfies a linearized iteration equation
of the form
\be
\tu_{n+1}=\left( 1+ \D \tau\,{\cal L} \right) \tu_n\,, \qquad \D \tau >0\,,
\label{e5_02}
\ee
where  $\cal L$ is the linear operator that results
when the iteration method is linearized on the background of the solitary wave $u$,
and $\D \tau$ is an auxiliary scaling parameter. From a conceptual point of view, 
the presence of $\D\tau$ emphasizes the analogy of iteration methods with numerical
methods of solving time-dependent differential equations (see, e.g., \cite{ChehabL05});
from a practical point of view, it can be used to ensure (in certain cases) or optimize
the convergence of the method, as we will discuss later on.

Let us begin with general remarks regarding the convergence rate of the linearized
iteration equation (\ref{e5_02}). Suppose that the
eigenfunctions of $\cal L$ form a complete set in an appropriate functional space, so that $\tu_n$ 
can be expanded over them. Let the minimum and maximum eigenvalues of $\cal L$ be $\Lambda_{\min}$ 
and $\Lambda_{\max}$. Then the convergence rate of the iteration method can be defined as \
log$(1/R)$, where the convergence factor $R$ is the maximum (in magnitude) eigenvalue of the
operator on the r.h.s. of (\ref{e5_02}):
\be
R=\max \left\{ |1+\Lambda_{\max}\D \tau|, \; |1+\Lambda_{\min}\D \tau| \right\}\,.
\label{e5_03}
\ee
Clearly, $R<1$ needs to hold in order for the iterations to converge, which implies
\be
\Lambda_{\max}\le 0 \quad {\rm and} \quad 1+\Lambda_{\min}\D \tau > -1\,.
\label{e5_04}
\ee
Moreover, if $\Lambda_{\max}=0$, then the corresponding
eigenfunction of $\cal L$ needs to be a translational
eigenmode (if it exists) of the linearized Eq. (\ref{e5_01}), which
only shifts the solution in space and hence does not affect the
convergence of the method. The smaller the convergence factor $R$,
the faster the convergence. It can be readily shown \cite{YangL06}
that the minimum value of $R$ occurs at
\be
\D \tau_*=\frac{-2}{\Lambda_{\min}+\Lambda_{\max}}
\label{e5_05}
\ee
(recall that $ \Lambda_{\min} < \Lambda_{\max} < 0$) and equals
\be
R_*= \frac{ 1- (\Lambda_{\max} / \Lambda_{\min} ) }{1+ (\Lambda_{\max} / \Lambda_{\min} ) } .
\label{e5_06}
\ee
Therefore, the closer the ratio $(\Lambda_{\max} / \Lambda_{\min} )$
to 1, the faster the convergence of the iteration method. 
Below we will compare the possible values of this ratio for the
generalized \PM\ and the accelerated ITEM. To that end, we first need
to cast the linearizations of these methods into the form of Eq. (\ref{e5_02}).

Let $L_0$ denote the nonlinear operator of Eq. (\ref{e5_01})
(or, more generally, of the stationary wave equation whose solution we are looking for), 
so that that equation is rewritten as
\be
L_0u=0\,.
\label{p2_5_02}
\ee
Let $L$ be the corresponding linearized operator, so that
\be
L_0(u+\tu) \equiv L_0u+L\tu = L\tu\,, \qquad \mbox{for any}\; |\tu|\ll |u|\,.
\label{p2_5_03}
\ee
Note that for Hamiltonian wave equations, $L$ is always self-adjoint.
With these notations, the generalized \PM\ is \cite{gP}:
\be
u_{n+1}-u_n = \left( N^{-1} (L_0u)_n - \gamma\,
 \frac{ \langle u_n, (L_0 u)_n \rangle }{ \langle u_n, N u_n \rangle }\,u_n \right)\D \tau\,,
\label{p2_5_04}
\ee
where
\be
\gamma=1+\frac1{\alpha \D \tau}\,.
\label{p2_5_05}
\ee
Here and below, the inner product between two real-valued functions is defined in a standard way:
$$
\langle f, g \rangle = \Int f(\vecx)g(\vecx)\,d\vecx\,.
$$
For the positive definite and self-adjoint operator $N$ in 
(\ref{p2_5_04}), we take the simplest form used in \cite{gP}:
\be
N=c-\nabla^2\,,
\label{p2_5_06}
\ee
where the constant $c$ is given by Eq. (3.11) of \cite{gP}. The constant $\alpha$ in
Eq. (\ref{p2_5_05}) above is such that
\be
Lu \approx \alpha Nu
\label{p2_5_07}
\ee
in a certain least-square sense; a formula for computing this constant at each iteration
can be found in either of Eqs. (3.12) or (3.15) of \cite{gP}, but will not be needed here.
Following the steps of a calculation found at the beginning of Section 2 of \cite{gP}, it
is straightforward to show that the linearized form of 
the generalized \PM\ is:
\be
\tu_{n+1}-\tu_n= \left( N^{-1}L \tu_n - \gamma
\frac{ \langle u, L\tu_n \rangle }{ \langle u, Nu \rangle }\,u \,\right) \D \tau\,.
\label{e5_07}
\ee

\begin{figure}
\rotatebox{0}{\resizebox{10cm}{13cm}{\includegraphics[0in,0.5in]
 [8in,10.5in]{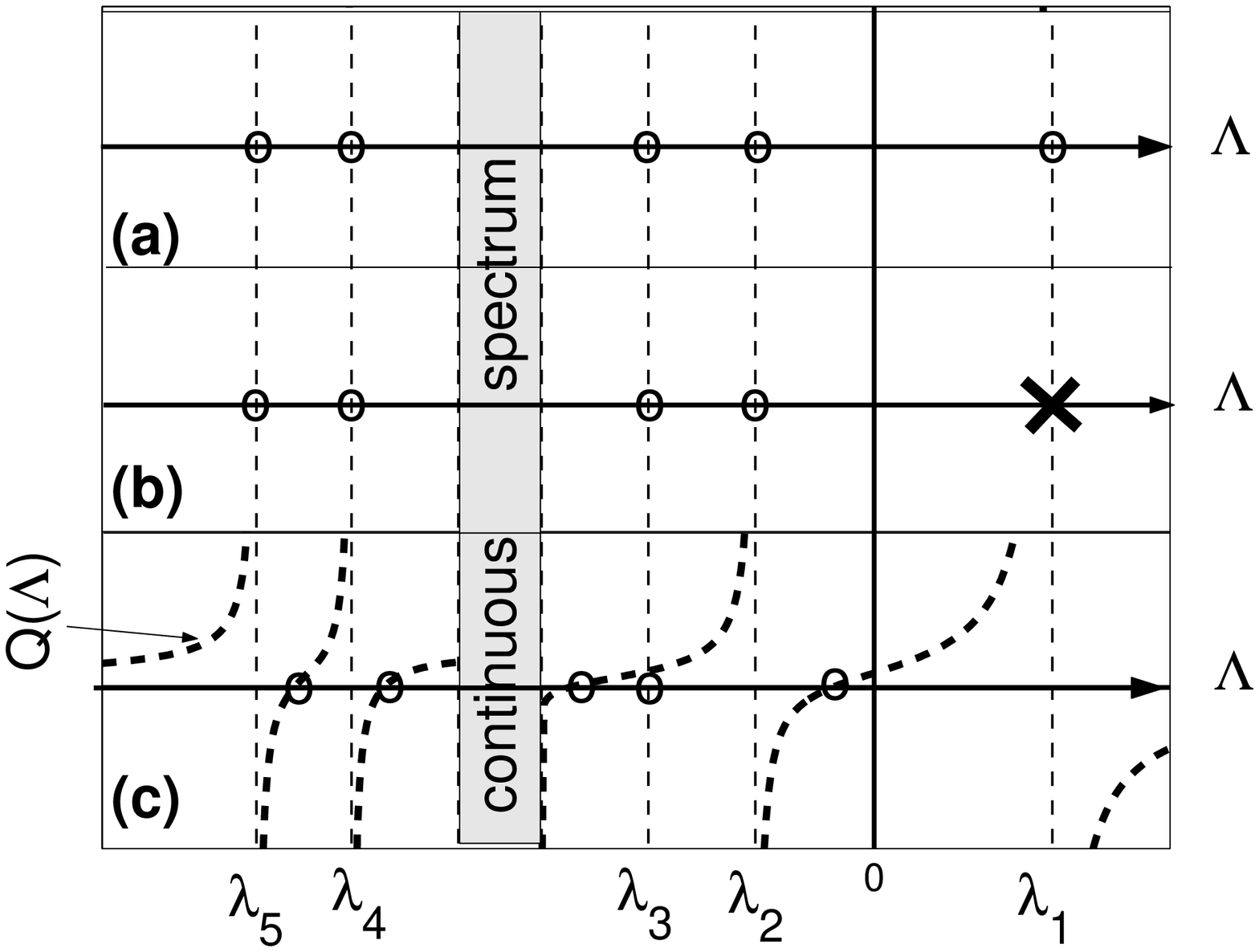}}}
\vspace{-2cm}
\caption{Schematics of the spectra of the operators found on the r.h.s.'es of (\ref{e5_11}) (a),
(\ref{e5_07}) (b), and (\ref{e5_10}) (c). The circles show the location of discrete eigenvalues. The
cross on the right of panel (b) indicates the disappearance of the eigenvalue compared to panel (a).
The thick dashed line in panel (c) shows a sample function of Eq. (\ref{e5_12}). The left edge
of the continuous spectrum is located at $\Lambda=-1$. It is assumed that $\langle u, \, \psi_3 \rangle=0$;
see text after (\ref{e5_12}).
}
\label{fig8}
\end{figure}

Next, the accelerated ITEM of Ref. \cite{YangL06} is:
\be
u_{n+1}=\left[ \frac{P}{ \langle \hat{u}_{n+1}, \hat{u}_{n+1} \rangle } \right]^{\frac12} \hat{u}_{n+1},
\label{add1_5_01}
\ee
\be
\hat{u}_{n+1}-u_n = K^{-1} \left( \nabla^2 u_n + F(\vecx,u_n)-\mu_n u_n \right)\D \tau,
\label{e5_08}
\ee
\be
\mu_n=\frac{ \langle \nabla^2 u_n + F(\vecx,u_n), K^{-1}u_n \rangle }
 { \langle u_n, K^{-1} u_n \rangle },
\label{e5_09}
\ee
where \ $P=\Int u^2 d\vecx$ \ is the specified power of the solitary wave. The 
positive definite and self-adjoint operator $K$ is referred to as the 
acceleration operator for the ITEM \cite{GarciaRipollPG01,YangL06}.
For simplicity, we take $K$ to have the same form (\ref{p2_5_06})
as the operator $N$ in the generalized \PM, with the $c$ being now an arbitrary positive constant.
The linearized form of ITEM (\ref{add1_5_01})--(\ref{e5_09}) is \cite{YangL06}:
\be
\tu_{n+1}-\tu_n = \left(  K^{-1}L \tu_n -
\frac{ \langle u, K^{-1} L\tu_n \rangle }{ \langle u, K^{-1} u \rangle }\,K^{-1} u  \right) \D \tau\,.
\label{e5_10}
\ee
Thus, the ``primordial" operator in the linearized equations of both the generalized \PM\
and the accelerated ITEM has the form:
\be
\hat{L}= (c-\nabla^2)^{-1}L\,.
\label{e5_11}
\ee
With $L$ being the linearized operator of (\ref{e5_01}), the
continuous spectrum of $\hat{L}$ is an interval (or, when
$F(\vecx,u)$ is a periodic function of $\vecx$, a union of
intervals), one of the end points of which is $\lambda=-1$ (see,
e.g., \cite{YangL06} and references therein). This eigenvalue of
$\hat{L}$ corresponds to the eigenvalue $\lambda=-\infty$ of $L$.
Then a possible spectrum of $\hat{L}$ is shown in
Fig.~\ref{fig8}a.

Even though the first terms on the r.h.s.'s of (\ref{e5_07}) and (\ref{e5_10}) 
have the same form (\ref{e5_11}), 
the eigenvalues of the corresponding operators $\cal L$ are 
different for two reasons. First, 
the values of $c$ in operators $N$ and $K$ are, in general, different, which makes
different the eigenvalues of the corresponding $\hat{L}$'s. Second,
the nonlocal terms (involving inner products) in
(\ref{e5_07}) and (\ref{e5_10})  modify the eigenvalues of $\hat{L}$ in different
ways. We now consider this latter issue in more detail.

In regards to the operator of the linearized \PM\ (\ref{e5_07}), 
we recall a fact \cite{gP} that is {\em important for our discussion} 
both here and in the next Section.
Namely, the role of the nonlocal term in 
that operator is to (nearly) eliminate from $\tu_{n+1}$ the eigenfunction of
$\hat{L}=N^{-1}L$ whose profile is close to that of the solitary
wave $u$, while leaving the other eigenfunctions and their eigenvalues
(nearly) unchanged.
This is ensured by taking the constant $\gamma$ and the operator $N$ 
to satisfy (\ref{p2_5_05}) and (\ref{p2_5_06}), respectively. 
The adverb ``nearly" is used above to account for the fact that 
relation (\ref{p2_5_07}) for Eq. (\ref{e5_01}) with a general nonlinear
function $F(\vecx,u)$ holds only approximately.
It is exact only for wave equations with power-law nonlinearity \cite{PelinovskyS04}, 
for which the original \PM\ was proposed \cite{Petviashvili76}.
However, the special choice of the constant $c$ in (\ref{p2_5_06}),
as noted after that equation, makes the approximation in (\ref{p2_5_07}) 
sufficiently accurate at least near the ``core" of the solitary wave.

Continuing with the discussion about the effect of the nonlocal term in 
(\ref{e5_07}) on the eigenvalues of the corresponding operator $\cal{L}$, 
let us suppose that $u$ is a {\em fundamental} solution 
of the nonlinear wave equation. 
(E.g., in the case of
Eq. (\ref{e5_01}), the fundamental solution, unlike nonfundamental ones,
 has no nodes\footnote{By
nodes in $D>1$ spatial dimensions, we mean sets of points of
dimension less than $D$ where $u(\vecx)=0$.}.
For a more general Eq. (\ref{p2_1_01}) where the operator $M$ is different from $\nabla^2$, 
fundamental solutions may have nodes (as, e.g., the lump solution of the Kadomtsev-Petviashvili
equation \cite{LumpKP}); in that case, their distinguishing feature
is that they have one ``main" hump, while the nonfundamental
solutions usually have several ``main" humps.) 
Then the ``$u$-like" eigenfunction of operator $N^{-1}L$ mentioned in the previous paragraph
(see also (\ref{p2_5_07})) corresponds to the largest eigenvalue, $\lambda_1$,
of that operator; see Fig.~\ref{fig8}a. Since this eigenfunction is eliminated by the nonlocal term 
at each iteration, then the resulting spectrum of the operator on
the r.h.s. of (\ref{e5_07}) is as shown in Fig.~\ref{fig8}b.
 Thus, for this operator, $\Lambda_{\max} \approx \lambda_2$ and $\Lambda_{\min} \approx \lambda_{\min}$;
the reason for using ``$\approx$" instead of ``$=$" is that relation (\ref{p2_5_07}) holds
approximately, as we noted above.
Now, if $\lambda_2<0$ and the step size $\D \tau$ satisfies a condition 
\be
1+\lambda_{\min}\D \tau > -1\,,
\label{p2_5_08}
\ee
then according to (\ref{e5_04}),  the generalized \PM\
converges to $u$. As a sidenote, we mention that
 for equations with power-law nonlinearity, $L$ is known \cite{Weinstein}
to have only one positive eigenvalue, and hence the Sylvester inertia law (see, e.g., Theorem 7.6.3
in \cite{HornJohnson91}) guarantees that $\lambda_1$ is the only positive eigenvalue of
$\hat{L}=N^{-1}L$.

Now let us consider the linearized operator $\cal{L}$ in (\ref{e5_10}) for
the ITEM (\ref{add1_5_01})--(\ref{e5_09}). In \cite{YangL06},
we showed that the set of discrete eigenvalues of this $\cal{L}$ is
the union of two sets: \ (i) the roots of a function
\be
Q(\Lambda)=\sum_j \frac{ | \langle u,\, \psi_j \rangle |^2 }{ \lambda_j -\Lambda } \,+\,
 \int_{\rm continuum} \frac{ | \langle u,\, \psi(\lambda) \rangle |^2 \,d\lambda}{ \lambda -\Lambda } \,,
\label{e5_12}
\ee
where $\psi_j$ is the eigenfunction of $\hat{L}$ corresponding to the eigenvalue $\lambda_j$, \ and also
\ (ii) the set of those $\lambda_j$ for which $\langle u, \,\psi_j \rangle=0$. This is shown schematically
in Fig.~\ref{fig8}c, with $\psi_3$ there satisfying $\langle u, \,\psi_3 \rangle=0$. (Note that
$Q(\Lambda)$ does not need to be defined for the continuum eigenvalues $\Lambda$.) Thus, for the operator
$\cal{L}$ in (\ref{e5_10}), \ $\Lambda_{\min} \ge \lambda_{\min}$ and $\Lambda_{\max} > \lambda_2$.

The consideration of the two preceding paragraphs shows that even
when the acceleration operators $N$ and $K$ in (\ref{e5_07}) and
(\ref{e5_10}) are the same (i.e., have the same $c$), one cannot, in
general, make a definite statement on whether the ratio
$(\Lambda_{\max}/\Lambda_{\min})$, and hence the convergence rate,
is greater for the generalized Petviashvili method or for the accelerated ITEM.
Moreover, the fact that the values of $c$ in $N$ and $K$ are
generally different, and hence so are the eigenvalues $\lambda_j$ of the corresponding 
two $\hat{L}$'s,  further obstructs the comparison of the
convergence rates of the two methods. The only two statements
that can be made here are the following. \ (i) For equations (\ref{e5_01})
with arbitrary nonlinearity, if the ITEM converges
to a fundamental solution, 
then we expect that in most cases (see below), so does the generalized \PM. \ 
(ii) For equations (\ref{e5_01}) with power-law nonlinearity $F(\vecx,u)=u^p$,
the \PM\ with the optimal choice of $\D \tau$ 
converges to the fundamental solution faster 
than does the optimally accelerated ITEM (\ref{add1_5_01})--(\ref{e5_09}).

To justify statement (i), first recall that for fundamental solitary waves,
\be 
(\Lambda_{\max})_{\rm Petviashvili} \approx \lambda_2 <
(\Lambda_{\max})_{\rm ITEM}\,,
\label{e5_13} 
\ee
as long as the value $c$ in the operator (\ref{e5_11}) is taken to be the 
same for both methods. Next, if the ITEM converges, then according to (\ref{e5_04}),
$(\Lambda_{\max})_{\rm ITEM}<0$, thereby implying that $\lambda_2<0$. 
However, by the Sylvester inertia law, the sign of $\lambda_2$ does not
depend on the actual value of $c$ (as long as $c>0$). Therefore, with a possible exception 
of those cases where $\lambda_2$ is close to zero,  
the left part of (\ref{e5_13}) yields $(\Lambda_{\max})_{\rm Petviashvili} < 0$,
which means that the generalized \PM\ converges. 
To prove statement (ii), first note that
operator $L$ in this case satisfies the conditions of Theorem 4 of Ref. \cite{YangL06}, 
so that $c=\mu$ is the optimal value for $K$ and  
$(\Lambda_{\min})_{\rm ITEM}=\lambda_{\min} \;(=-1)$. Next, 
in the \PM\ for the equation with $F(\vecx,u)=u^p$, $N=M$ \cite{Petviashvili76,gP}
and hence $c=\mu$ as well, whence 
$(\Lambda_{\min})_{\rm Petviashvili} = \lambda_{\min}$. Thus, in this case,
\be
(\Lambda_{\min})_{\rm Petviashvili} = (\Lambda_{\min})_{\rm ITEM}.
\label{e5_14}
\ee
Combining Eq.~(\ref{e5_14}) and inequality (\ref{e5_13}), where now the sign ``$\approx$" must be
replaced with ``$=$", one concludes that $(\Lambda_{\max}/\Lambda_{\min})$ should be greater for the \PM;
hence statement (ii) follows.

A simple example illustrating statement (ii)
is the stationary nonlinear Schr\"odinger equation in one dimension:
\be
u_{xx}+u^3=u, \qquad |u|\To 0 \quad {\rm as} \quad |x|\To\infty,
\label{e5_15}
\ee
for which the ITEM with the parameters $c=\mu\,(=1)$ and $\D
\tau = 1.5 $, corresponding to the optimal acceleration, converges to the accuracy of $10^{-10}$ in 33
iterations. The \PM\ (\ref{p2_5_04}) with $\D \tau=1.5$, $alpha=2$ (as in the original \PM; 
see \cite{gP}), and $\gamma$ given by (\ref{p2_5_05}), converges to the same accuracy in
19 iterations. Here both methods start with the initial condition
$u_0=e^{-x^2}$. In our numerical experiments of finding the fundamental solutions of non-power-law
equations (not covered by the above statement (ii)), 
we also observed that the
generalized \PM\ is faster than the optimally accelerated ITEM  (\ref{add1_5_01})--(\ref{e5_09});
see, e.g., Example 3.1 in \cite{gP}. (The ITEM with amplitude normalization \cite{YangL06} can still be
faster than the generalized \PM.)

\begin{figure}
\hspace*{0cm}
\mbox{
\begin{minipage}{7.5cm}
\rotatebox{0}{\resizebox{7.5cm}{10.4cm}{\includegraphics[0in,0.5in]
 [8in,10.5in]{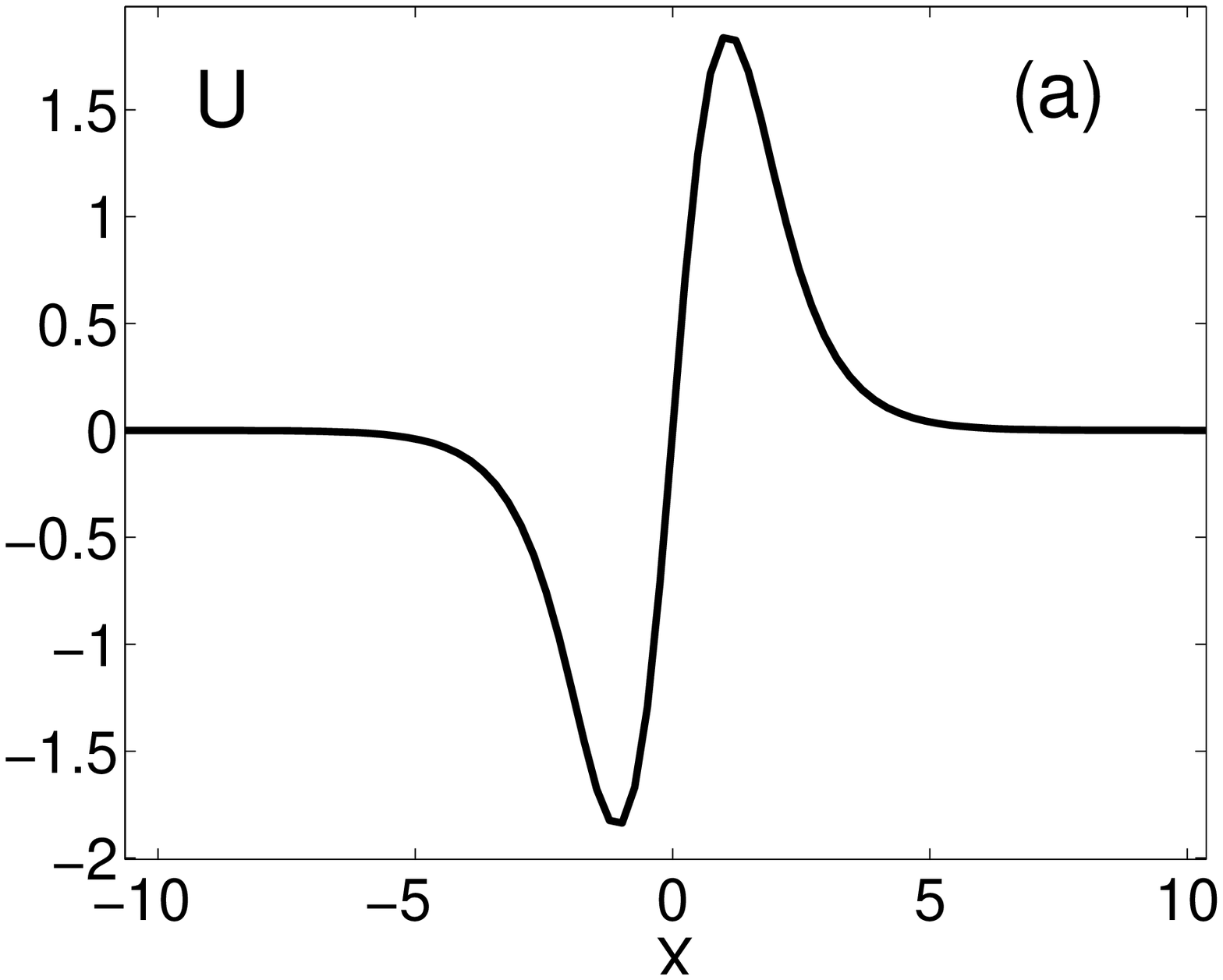}}}
\end{minipage}
\hspace{0.2cm}
\begin{minipage}{7.5cm}
\rotatebox{0}{\resizebox{7.5cm}{10.4cm}{\includegraphics[0in,0.5in]
 [8in,10.5in]{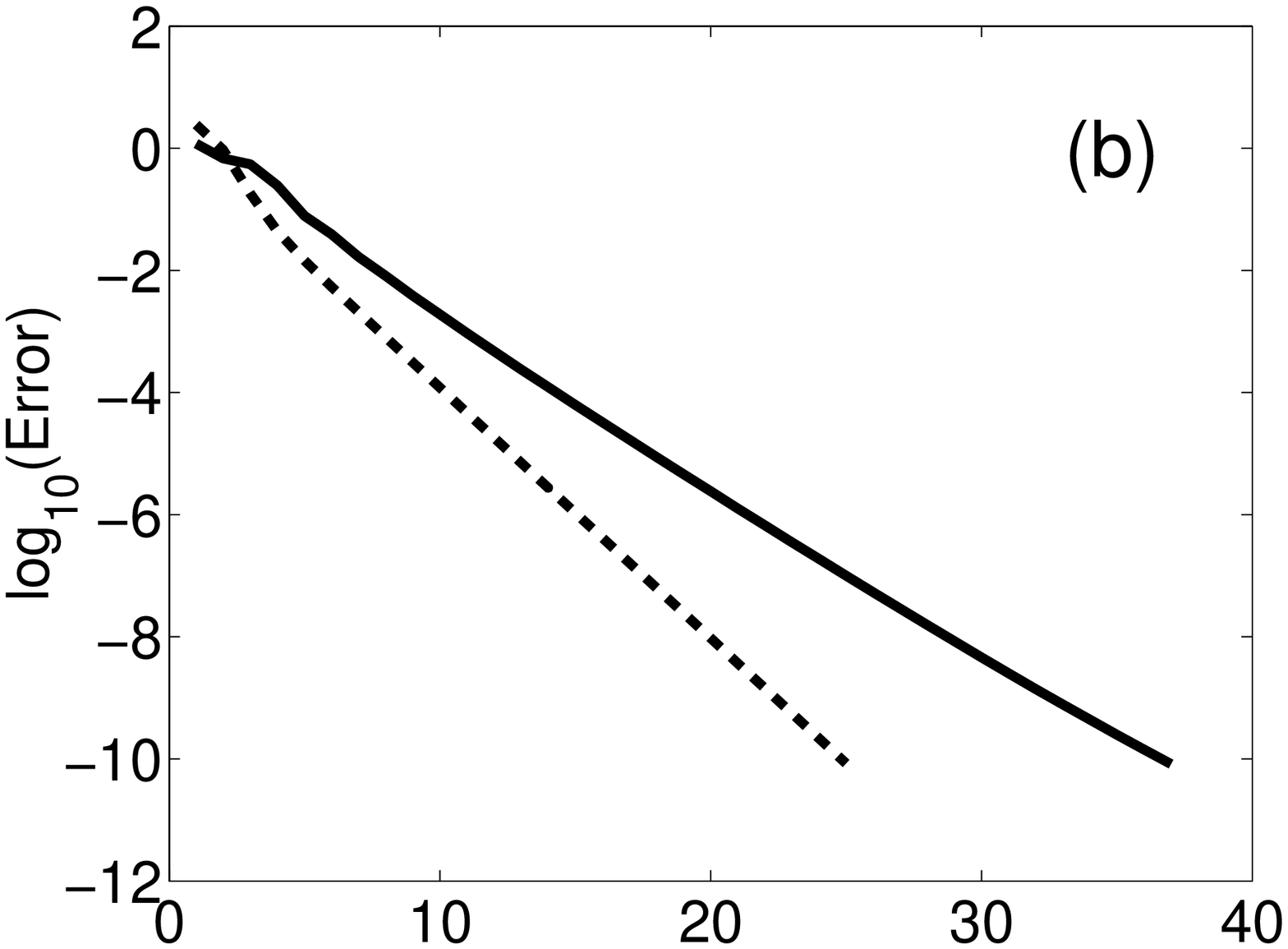}}}
\end{minipage}
 }
\vspace{-2.5cm}
\caption{ (a): The anti-symmetric solution of Eq. (\ref{p2_5_09}) with $\mu=1.43$ ($P=10$).
 \ (b): Error evolutions, starting with an anti-symmetric initial condition, 
for the generalized \PM\ (solid line) the optimally accelerated ITEM (dashed line).
}
\label{fig2}
\end{figure}

\begin{figure}
\rotatebox{0}{\resizebox{10cm}{13cm}{\includegraphics[0in,0.5in]
 [8in,10.5in]{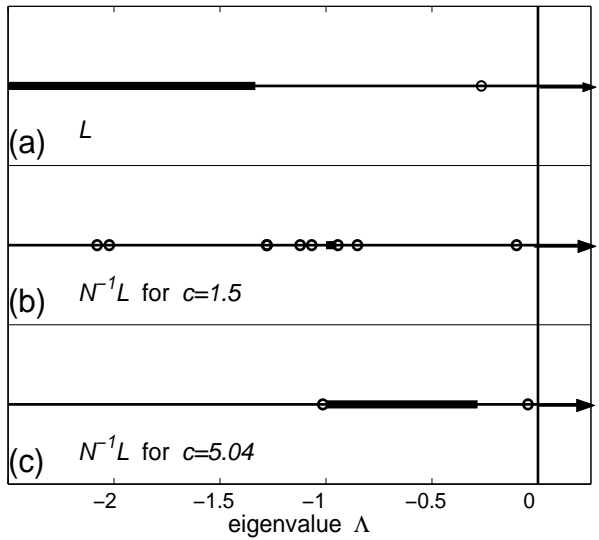}}}
\vspace{-3cm}
\caption{ The actual spectra of the linearized operator $L$ of Eq.~(\ref{p2_5_09}) (a) and of
the operators $N^{-1}L$, where $N$ is given by (\ref{p2_5_06}), with $c=1.5$ (b) and $c=5.04$ (c).
The operator in (b) has a very short interval of continuous spectrum between $-1$ and $-0.95$.
}
\label{fig9ADDED}
\end{figure}

However, in a situation where both methods converge to a
{\em nonfundamental} solitary wave, the
optimally accelerated ITEM can be faster than the generalized \PM.
As an example, let us revisit the equation with a double-well potential:
\be
u_{xx}+ V(x)u - u^3 = \mu u , \qquad V(x)=6\left( {\rm sech}^2(x-1) + {\rm sech}^2(x+1) \right),
\label{p2_5_09}
\ee
considered in Example 3.2 of \cite{gP}. We will focus on its anti-symmetric solution
(see Fig.~\ref{fig2}a) with the propagation constant $\mu=1.43$ and the corresponding 
power $P\equiv \Int u^2\,dx=10$.
This solution is nonfundamental since it has a node; the fundamental solution in
this case is a two-humped pulse with its maxima located near the
maxima of the potential.
The solid and dashed lines in Fig.~\ref{fig2}b show the evolutions of the error norm,
defined as
\be
E_n= \left( \frac{ \langle u_n-u_{n-1}, u_n-u_{n-1} \rangle}{ \langle u_n, u_n \rangle }
  \right)^{1/2}\,,
\label{p2_5_10}
\ee
for the generalized \PM\ and the optimally accelerated ITEM, respectively. 
In both cases, the parameter $\D\tau$ was emprically optimized (see (\ref{e5_05}))
to yield the maximum convergence rates; the respective values are
$\D\tau_{*\,,\rm Petviashvili}=1.6$ and $\D\tau_{*,\,\rm ITEM}=0.7$.
Also, in the case of the generalized \PM, the value $c=5.04$ was algorithmically
computed \cite{gP}, while for the ITEM, $c=1.5$ was empirically found to
yield the optimal convergence rate. 
As the initial condition for both these methods, we took \ $u_0=2x\,e^{-x^2}$.
As seen from Fig.~\ref{fig2}b, the optimally accelerated ITEM
is about one and a half times
faster than the generalized \PM.
The reason behind this can be understood by looking at the
spectra of the corresponding operators $\hat{L}$ in (\ref{e5_11}) with the above
values $c=1.5$ for the accelerated ITEM
(Fig.~\ref{fig9ADDED}b) and $c=5.04$ for the
generalized \PM\ (Fig.~\ref{fig9ADDED}c). Namely, when one starts
with an anti-symmetric initial condition (as we did above), 
the symmetric eigenmodes corresponding to
$\lambda_{2k+1}, \; k=0,1,\ldots$ do not contribute to the error
$\tu_n$. Then from (\ref{e5_06}) and Figs.~\ref{fig9ADDED}b,c,
$$
R_{\rm \;ITEM} < \frac{ 1 - (\lambda_{2,\;c=1.5} /\lambda_{\min,\;c=1.5})}{
  1 - (\lambda_{2,\;c=1.5} /\lambda_{\min,\;c=1.5}) } = \frac{1 - 0.41}{1+0.41}=0.42,
$$
$$
R_{\rm \;Petviashvili} \approx \frac{ 1 - (\lambda_{ \max\;{\rm continuum},\;c=5 }/\lambda_{\min,\;c=5}) }{
 1 + (\lambda_{ \max\;{\rm continuum},\;c=5 }/\lambda_{\min,\;c=5}) } = \frac{1 - 0.28}{1+0.28}=0.56,
$$
and hence the corresponding numbers of iterations to reach the accuracy of $10^{-10}$
can be estimated as:
$$
n_{\max,\;\rm ITEM} \approx \frac{ -10\ln 10}{\ln R_{\rm \;ITEM} } = 26, \qquad
n_{\max,\;\rm Petviashvili} \approx \frac{ -10\ln 10}{\ln R_{\rm \;Petviashvili} } = 40.
$$
These estimates are in very good agreement with the numbers of
iterations (25 and 37, respectively) reported in Fig.~\ref{fig2}b.
Note also that the empirically found optimal values of $\D\tau_*$
reported above agree with Eq.~(\ref{e5_05}) and the spectra shown in
Figs.~\ref{fig9ADDED}b,c. 


\setcounter{equation}{0}
\section{Mode elimination technique for improving convergence of iteration methods}

Here we develop the ideas of Ref. \cite{gP} and extend the
generalized \PM\ so that it could be employed for two additional purposes: \
(i) obtaining certain nonfundamental solutions of stationary
nonlinear wave equations; \ and \ (ii) accelerating convergence of
iterations methods. We emphasize that the technique we propose can
be applied to {\em any} iteration method and to single and
coupled equations as well. For simplicity of the presentation, below we
illustrate it for single equations of the form (\ref{e5_01}).

We begin with the observation that in most cases (with Eq. (\ref{p2_5_09})
being a notable exception), the generalized \PM\ would not converge
to a nonfundamental solution of a given wave equation. 
The reason for that can be understood from the following
simple example. Consider an equation
\be
u_{xx}+(6{\rm sech}^2x+u^2)u=\mu u\,.
\label{e6_01}
\ee
When the amplitude of $u$ is small, (\ref{e6_01}) has two solutions:
the fundamental, \ $\{ u^{(1)}\approx \epsilon\,{\rm sech}^2 x, \;
\mu^{(1)}\approx 4 \, \}$, \ and the nonfundamental, $\{
u^{(2)}\approx \epsilon\,{\rm sech} x \,{\rm tanh} x, \;
\mu^{(2)}\approx 1 \, \}$, where \ $\epsilon\ll 1$. Then the
operator obtained by the linearization of Eq. (\ref{e6_01}) on the
background of the nonfundamental solution,
\be
L \approx \partial_x^2 + 6{\rm sech}^2x - \mu^{(2)},
\label{e6_02}
\ee
has two largest eigenvalues: $\lambda_1\approx \mu^{(1)}-\mu^{(2)}\approx 3>0$ and
$\lambda_2\approx \mu^{(2)}-\mu^{(2)}=0$, with the corresponding eigenfunctions being approximately
$u^{(1)}$ and $u^{(2)}$. As we noted in Section 2, the nonlocal term in the linearized iteration equation
(\ref{e5_07}) nearly eliminates the eigenfunction of operator $\hat{L}=N^{-1}L$ which is ``similar" to the
background solution $u^{(2)}$. However, the eigenfunction of $\hat{L}$ corresponding to the eigenvalue
$\lambda_1>0$ of $\hat{L}$ is not eliminated, and hence, according to the discussion found before
Eq. (\ref{p2_5_08}), the generalized \PM\ will not converge to solution $u^{(2)}$.

The above example suggests a simple way in which the generalized
\PM\ (\ref{p2_5_04}) can be modified so that it would converge to a
nonfundamental solution $u$ (given, of course, an initial condition
close to $u$). In the general form, this modified method is 
\be
u_{n+1}-u_n= \left[ N^{-1} (L_0u)_n -
 \gamma\,\frac{ \langle u_n, \, (L_0u)_n \rangle }{\langle u_n, \, Nu_n \rangle }\,u_n -
 \sum_{j=1}^{J_{\rm unst}} \gamma_{\rm unst}^{(j)} \,
  \frac{ \langle \phi_{\rm unst}^{(j)} , \, (L_0u)_n \rangle }
  {\langle \phi_{\rm unst}^{(j)} , \, N \phi_{\rm unst}^{(j)} \rangle } \,\phi_{\rm unst}^{(j)}
 \right] \D \tau,
\label{e6_03}
\ee
where $\gamma$ and $N$ are defined as in (\ref{p2_5_05}) and (\ref{p2_5_06}), 
$\phi_{\rm unst}^{(j)}$ are the functions that approximate the eigenmodes of
operator $(N^{-1}L)$ with positive eigenvalues (excluding the
background solution $u$), $J_{\rm unst}$ is the number of such
eigenmodes, and 
\be
\gamma_{\rm unst}^{(j)} = 1+ \frac1{\alpha_{\rm unst}^{(j)} \D \tau},  \qquad
\alpha_{\rm unst}^{(j)} = \frac{ \langle \phi_{\rm unst}^{(j)} , \, L \phi_{\rm unst}^{(j)} \rangle }
  {\langle \phi_{\rm unst}^{(j)} , \, N \phi_{\rm unst}^{(j)} \rangle } \,.
\label{e6_04}
\ee
Here $\alpha_{\rm unst}^{(j)}$, defined analogously to (\ref{p2_5_07}):
\be
L \phi_{\rm unst}^{(j)} \approx   \alpha_{\rm unst}^{(j)} N \phi_{\rm unst}^{(j)}\,,
\label{p2_6_01}
\ee
is computed according to Eq. (3.12) of \cite{gP}.
In the context of the example in the previous paragraph, $J_{\rm unst}=1$ and
$ \phi_{\rm unst}^{(1)} = u^{(1)}$.

Following the lines of the analysis of Section 2 in Ref. \cite{gP}, it is
straightforward to show that in method (\ref{e6_03}), (\ref{e6_04}),
the components of the error $\tu_n$ ``aligned along" the modes
$\phi_{\rm unst}^{(j)}$, $j=1,\ldots, J_{\rm unst}$, are nearly
eliminated at every iteration; this is guaranteed by the form of the coefficients
 $\gamma_{\rm unst}^{(j)}$. 
Therefore, in what follows, we refer
to method (\ref{e6_03}) as the mode elimination method. In Section
4 below, we will present the results of applying this method to a
two-dimensional equation of the form (\ref{e6_01}) to obtain its
nonfundamental solutions. 

{\bf Remark} \ It is clear that the success of the mode
elimination method hinges upon the knowledge of the ``unstable"
eigenmodes $ \phi_{\rm unst}^{(j)}$. However, in many cases, an approximate knowledge of $\phi_{\rm
unst}^{(j)} $ may suffice. 

\smallskip

We now show how the mode elimination technique can be used to
accelerate convergence of iteration methods. 
The reason that a given method converges slowly is, 
according to (\ref{e5_06}), that the ratio $\Lambda_{\max}/\Lambda_{\min}$
is small. Since for an appropriately chosen operator $N$,
$|\Lambda_{\min}|=O(1)$ (see, e.g., Figs.~\ref{fig9ADDED}b,c), then
for a slowly convergent method, the eigenvalue $|\Lambda_{\max}|$ must be
small. Then if one can eliminate the
corresponding eigenmode, similarly to how it is 
done in (\ref{e6_03}), one essentially replaces $(\Lambda_{\max})_{\rm
old}$ with $(\Lambda_{\max})_{\rm new}<(\Lambda_{\max})_{\rm
old}\;(<0)$. Then the ratio $\Lambda_{\max}/\Lambda_{\min}$
increases and so does the convergence rate of the iteration method.
The practical issue here is how to find the mode, $\phi_{\rm slow}$,
which slows down the convergence. Fortunately, this is rather easy
to do using the following observation. For $\D \tau<\D \tau_*$,
where $\D \tau_*$ is defined in (\ref{e5_05}), the factor \ $
(1+\Lambda_{\rm slow}\D \tau) \equiv (1+\Lambda_{\max}\D \tau) $,
which governs the decay of $\phi_{\rm slow}$, is the largest among
such factors for all the eigenmodes of $(N^{-1}L)$. Then after some
iterations, the content of the error $\tu_n\equiv u_n-u$ becomes
dominated by the eigenmode $\phi_{\rm slow}$, and hence
\be
\phi_{\rm slow} \propto (u_n-u_{n-1})\,.
\label{e6_06}
\ee
The elimination of the function $(u_n-u_{n-1})$ is carried out in exactly the same way as in
(\ref{e6_03}), yielding the method:
\be
u_{n+1}-u_n= \left[ N^{-1} (L_0u)_n -
 \gamma\,\frac{ \langle u_n, \, (L_0u)_n \rangle }{\langle u_n, \, Nu_n \rangle }\,u_n -
 \gamma_{{\rm slow},\,n}\,\frac{ \langle \phi_{{\rm slow},\,n} , \, (L_0u)_n \rangle }
 {\langle \phi_{{\rm slow},\,n} , \, N \phi_{{\rm slow},\,n}  \rangle }\, \phi_{{\rm slow},\,n}
 \right] \D \tau,
\label{e6_07}
\ee
where
\be
\phi_{{\rm slow},\,n} = u_n-u_{n-1}, \qquad
\gamma_{{\rm slow},\,n} = 1+ \frac{s}{\alpha_{{\rm slow},\,n} \D \tau},  \qquad
\alpha_{{\rm slow},\,n} = \frac{ \langle \phi_{{\rm slow},\,n} , \, L \phi_{{\rm slow},\,n} \rangle }
  {\langle \phi_{{\rm slow},\,n} , \, N \phi_{{\rm slow},\,n} \rangle } \,.
\label{e6_08}
\ee
Note the coefficient $s$ in (\ref{e6_08}), which we will comment on
in the next paragraph. We will also provide examples that
demonstrate the efficiency of the accelerated Petviashvili method
(\ref{e6_07}), (\ref{e6_08}) and its extensions to other iteration
methods, in the next Section. 

Similarly to the analysis of Ref. \cite{gP}, one can show that
the role of coefficient $s$
in (\ref{e6_08}) is to control how much of the mode $ \phi_{{\rm
slow},\,n}$ is subtracted at each iteration. We found empirically
that in most cases, it is beneficial for the convergence rate to
subtract not the entire $\phi_{{\rm slow},\,n}$-component from $u_n$
but only part of it, usually somewhere between 40\% and 80\% (i.e.,
use $s\sim 0.4 \mbox{--} 0.8$). 
(However, even using the value $s=1$
leads to a significant increase in convergence rate compared to the
corresponding non-accelerated method  when the latter is slow.)
The justification of using $0<s<1$ (or, alternatively, $1<s<2$)
rather than $s=1$, is based on the same considerations, found before Eq.
(\ref{e6_06}), which led us to propose the accelerated method
(\ref{e6_07}). Namely, to uphold those considerations, $ \phi_{{\rm
slow},\,n} $ is to remain the most slowly decaying eigenmode of
$(N^{-1}L)$ at every iteration. In the case where the entire amount
of it is subtracted at the $(n+1)$st iteration, it is not obvious
(and probably not true) that the error $\tu_{n+2}$ at the next iteration
would consist mainly of the mode $\phi_{{\rm slow},\,n+1}\equiv
u_{n+1}-u_n$, which will be subtracted at the $(n+2)$nd
iteration. However, if only $s\cdot 100$\% of mode $ \phi_{{\rm
slow},\,n} $ is subtracted, this mode can still remain the most
slowly decaying as long as
\be
| (1-s)\cdot(1+ \Lambda_{\rm slow} \D \tau ) | > |1+ \Lambda_{\rm next} \D \tau|,
\label{e6_09}
\ee
where $ \Lambda_{\rm slow} $ is the eigenvalue corresponding to $
\phi_{{\rm slow},\,n} $, and $ \Lambda_{\rm next} $ is the
eigenvalue corresponding to the next most slowly decaying mode. Yet,
for $s$ not too small, the l.h.s. of (\ref{e6_09}) is still
considerably less than $| 1+ \Lambda_{\rm slow}\D \tau |$, and hence
the convergence rate of the original iteration method is increased.

To conclude this Section, we compare our mode elimination
technique for convergence acceleration with the Steffensen's method
(see, e.g., \cite{NA}), which is based on applying the Aitken's
acceleration algorithm every given number of iterations.
The idea of the Steffensen's method is the following. Suppose one
has three consecutive iterative solutions $u_n$, $u_{n+1}$,
$u_{n+2}$ about which one knows that they satisfy
\be
\frac{ \tu_{n+2} (\vecx) }{ \tu_{n+1} (\vecx) } \approx
\frac{ \tu_{n+1}(\vecx)  }{ \tu_{n}(\vecx)  } \qquad \mbox{for all $\vecx$},
\label{e6_10}
\ee
where $\tu_n$ is the error defined in (\ref{p2_5_01}). Using these solutions, one applies the
Aitken's algorithm:
\be
u_{n+3}\equiv u_n^A = u_n - \frac{(u_{n+1}-u_n)^2}{u_{n+2}-2u_{n+1}+u_n}\,,
\label{e6_11}
\ee
and then proceeds to computing the next few iterations $u_{n+4}, \,\ldots\,, u_{n+n_{\rm accel}+2}$ 
with the original iteration method, where $n_{\rm accel}\ge 3$. 
Then one uses $u_{n+n_{\rm accel}}$, $u_{n+n_{\rm accel}+1}$,
$u_{n+n_{\rm accel}+2}$ to compute $u_{n+n_{\rm accel}}^A$ by (\ref{e6_11}) with $n\To n+n_{\rm accel}$,
and so on. In \cite{DemanetS06}, this method was successfully used to accelerate the convergence of the
original \PM\ for the nonlinear Schr\"odinger equation in 3 spatial dimensions.

Aitken's algorithm (\ref{e6_11}) systematically reduces the error
$(u_n^A-u)$ only when (\ref{e6_10}) holds sufficiently well, which
occurs under the same condition (\ref{e6_06}) that must hold in
order for the mode elimination method to work. However, the {\em
sense} in which (\ref{e6_06}) is to hold is drastically
different for these two acceleration techniques. For the mode
elimination, it suffices if (\ref{e6_06}) holds approximately {\em near
the ``core"} of the solitary wave, since $\phi_{\rm slow}$ enters 
Eqs. (\ref{e6_07}), (\ref{e6_08}) via the inner products with
functions that are essentially nonzero only in that spatial region.
On the contrary, for the
Steffensen's method, (\ref{e6_10}) has to hold {\em pointwise} and,
in particular, far away from the ``core" of $u(\vecx)$. In the latter
spatial region, the denominator of (\ref{e6_11}) is nearly zero, and
hence even a small ripple in $u_n$, $u_{n+1}$, or $u_{n+2}$ can
result in a large distortion of $u_n^A$. This was indeed observed in
our numerical experiments, except in the cases where 
$ |\Lambda_{\max}| \ll |\Lambda_{\rm next}| $,
where $|\Lambda_{\rm next}|$ is defined after (\ref{e6_09}). Thus, we expect
our mode elimination technique and the Steffensen's method to be
competitive in those latter cases, but expect the mode elimination
technique to have superior performance over that of the Steffensen's
method when there are more than one eigenmodes with $\Lambda \approx \Lambda_{\max}$.
This expectation is borne out by Examples 4.2 and 4.3 reported below.


\setcounter{equation}{0}
\section{Examples of the mode elimination technique}

Below we illustrate the application of the mode elimination technique 
to obtaining nonfundamental solitary waves and to accelerating convergence
of iteration methods for Eq. (\ref{e5_01}).
In Ref. \cite{SO}, we already showed by extensive simulations that
this technique can greatly accelerate convergence of a
class of universally-convergent iteration methods for both single
and coupled equations. (Method (\ref{e6_15}) presented below is a
particular member of that class.) Therefore, here we will
focus on clarifying the role of parameter $s$ in Eq. (\ref{e6_08}) 
for optimizing the convergence rate 
and also on demonstrating the applicability of the mode elimination
technique to {\em various} classes of iteration methods.

\smallskip

{\bf Example 4.1} \ Here we will demonstrate that method
(\ref{e6_03}), (\ref{e6_04}) can be used to obtain nonfundamental
solitary waves when approximate information about the unstable
eigenmodes of $(N^{-1}L)$ is available. We will also compare the
performance of this method with that of a universally-convergent
method proposed in \cite{SO}.

\begin{figure}
\vspace*{-7cm}
\rotatebox{0}{\resizebox{10cm}{13cm}{\includegraphics[0in,0.5in]
 [8in,10.5in]{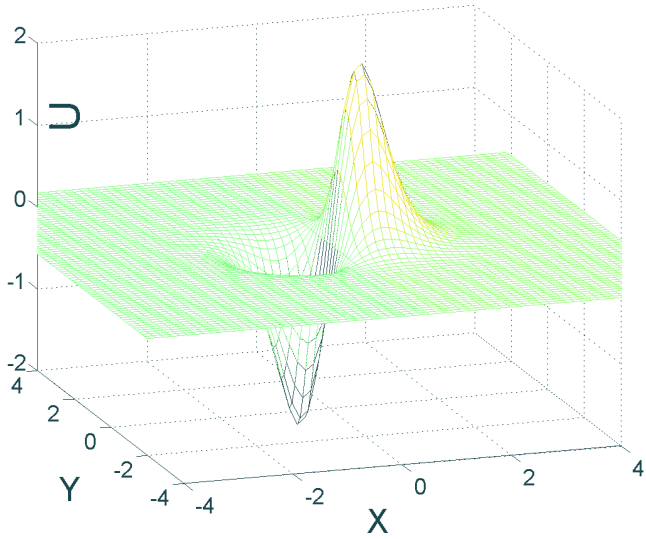}}}
\vspace{1cm} \caption{The first nonfundamental solution of Eq.
(\ref{e6_12}) with $\mu=8$. } 
\label{fig9}
\end{figure}

Equation
\be
\nabla^2 u + V_0({\rm sech} \,x\,{\rm sech}\, y)^2 u + u^3 = \mu u, \qquad V_0=20
\label{e6_12}
\ee
is a two-dimensional counterpart of Eq. (\ref{e6_01}). Since the
potential well in (\ref{e6_12}) is sufficiently deep ($V_0\gg 1$),
this equation admits several nonfundamental solutions. Below we
report the details of finding the first of them which corresponds to
$\mu=8$ and is shown in Fig.~\ref{fig9}. For this 
solution, we expect the generalized Petviashvili method to have one
unstable eigenmode (in addition to the mode approximated by $u$ that may possibly also be 
unstable), and approximate this eigenmode by
\begin{equation}
\phi_{\rm unst}= e^{-\frac12 (r/W)^2} ,  \qquad r^2=x^2+y^2\,.
\label{e6_13}
\end{equation}
The width $W$ in (\ref{e6_13}) is found iteratively from the formula
\be 
W_n^2= \frac23 \, \frac{ \langle u_n, \, x^2 u_n \rangle }{
\langle u_n, \, u_n \rangle },
\label{e6_14} 
\ee
in deriving which we assumed that $u \propto x \,\phi_{\rm unst}$.
Starting with the initial
condition $u_0=2x\,e^{-(x^2+y^2)}$, method (\ref{e6_03}),
(\ref{e6_04}) with a nearly optimal $\D \tau=0.7$ took about 50
iterations to reach the accuracy of $10^{-10}$. Thus, the
generalized Petviashvili method with mode elimination (\ref{e6_03}),
(\ref{e6_04}) converges to this nonfundamental solution, while the
generalized Petviashvili method (\ref{p2_5_04}) without the mode elimination diverges.

We also obtained the same solution by a method based on the ``squared" operator $(N^{-1}L)$:
\be
u_{n+1}-u_n= - \left[ (N^{-1}LN^{-1}L_0 u)_n - \Gamma_n
  \frac{ \langle u_n, \, (LN^{-1}L_0 u)_n \rangle }{ \langle u_n, \, Nu_n \rangle }\, u_n \,\right] \D \tau.
\label{e6_15}
\ee
(The name ``squared" comes from the fact that $(N^{-1}L)^2$ appears 
in the linearized version of (\ref{e6_15}).)
In \cite{SO}, we showed that this method belongs to a family of
universally-convergent methods (i.e., methods which can converge to
{\em any} nonfundamental solution of a given equation provided that
the initial condition is sufficiently close to that solution) for
either of the following choices of $\Gamma_n$: \ $\Gamma_n=0$ or
\be
\Gamma_n=1 - \frac1{ \left( \langle u_n, \, (LN^{-1}L u)_n \rangle \,/\,
 \langle u_n, \, Nu_n \rangle \right)\, \D \tau } \,.
\label{e6_16}
\ee
Note that this $\Gamma_n$ is defined similarly to $\gamma_n$ in the
generalized \PM\ (see Eqs. (\ref{p2_5_05}) and (\ref{e6_04})).
 Since we are looking for a nonfundamental solution
of (\ref{e6_12}), then using the value for $\Gamma_n$ given by
(\ref{e6_16}) as opposed to $\Gamma_n=0$ will not eliminate the mode
with the maximum eigenvalue (see the discussion after Eq. (\ref{e6_02})),
 and hence will not speed up the convergence of the
iterations. Therefore, in the remainder of this Example we report
the results for method (\ref{e6_15}) with $\Gamma_n=0$. 
Starting with the same initial condition as above, this method with
the operator $N$ computed as in \cite{gP} and with a 
nearly optimal $\D \tau =0.5$ took about 190 iterations to converge to
the accuracy of $10^{-10}$. Thus, the mode elimination method
(\ref{e6_03}), (\ref{e6_04}) is several times faster than the
squared-operator method (\ref{e6_15}) for finding the first
nonfundamental solution of (\ref{e6_12}). (We also observed that
method (\ref{e6_03}), (\ref{e6_04}) is less sensitive to the choice
of initial conditions than method (\ref{e6_15}).) However, when we
additionally included the step of eliminating the slow mode, as in
Eqs.~(\ref{e6_06})--(\ref{e6_08}), into both methods, the difference
in their convergence rates was significantly reduced. Namely, the
convergence of method (\ref{e6_03}), (\ref{e6_04}), which has
already been quite rapid, was not improved by this additional step
(and the number of iterations remained around 50), while the
squared-operator method now took about 70 iterations to converge.

We also applied both methods to finding the second nonfundamental
solution of (\ref{e6_12}), which has the shape similar to
$A(1-Br^2)\,e^{-(r/C)^2}$ with $r^2=x^2+y^2$ and $A,B,C={\rm const}$
(see Fig.~\ref{fig12} below). For this solution we
found, through experimentation, that one needs to include five unstable modes into
(\ref{e6_03}). 
For the respective optimal $\D \tau$'s, the generalized Petviashvili method
with mode elimination (\ref{e6_03}), (\ref{e6_04}) was found to be
about 50\% faster than the squared-operator method (\ref{e6_15}). However, this
advantage in the convergence rate is offset by the increased
complexity arising from the need to guess the number and profiles of unstable modes
and then to estimate their parameters (namely, the widths). 
Therefore, we conclude that the mode
elimination method may be more efficient than the squared-operator
method for finding the {\em lowest-order} nonfundamental solitary waves,
as long as some reasonable guess about the unstable modes can be
made. However, for finding second- and higher-order nonfundamental
solutions, method (\ref{e6_15}) appears to be easier to implement
and hence more practical.

\smallskip

{\bf Example 4.2} \
In this and the next two Examples, we demonstrate the efficiency of the convergence acceleration technique
based on the mode elimination, as in (\ref{e6_07}) and (\ref{e6_08}), for three
different iteration methods. In this Example,
we apply this technique to the generalized \PM.

We look for the fundamental solitary wave of an equation arising in the theory of 
nonlinear photonic lattices:
\be
\nabla^2 u + V_0 (\cos^2x + \cos^2 y) u+u^3 = \mu u\,.
\label{p2_6_02}
\ee
for three choices of the potential amplitude and the propagation
constant:
$$
(a): \quad V_0=4, \; \mu=4.95; \qquad (b): \quad V_0=4, \; \mu=6.5; \qquad (c): \quad V_0=0, \; \mu=1\,.
$$
In case $(a)$, the propagation constant is close to
the edge of the continuous spectrum band, and the solitary wave occipies many ``sites" of the potential,
while in case $(b)$, the propagation constant is sufficiently far away from the band edge, and the solitary
wave is well localized. (The profiles of the corresponding solutions are similar to those of the
top and bottom solutions shown in Fig.~3 of \cite{gP}.)
Case $(c)$ is that of the nonlinear Schr\"odinger equation in two spatial
dimensions. In all cases, we apply three methods: the generalized \PM\ (\ref{p2_5_04})
without any acceleration, the same method with the Aitken's acceleration (\ref{e6_11}) performed after
every third iteration $(n_{\rm accel}=3)$, and the mode elimination method (\ref{e6_07}), (\ref{e6_08})
with various values of $s$ (see the paragraph including Eq. (\ref{e6_09})). The initial condition in all
cases is \ $u_0=1.5\,e^{-(x^2+y^2)}$, 
and the step size $\D \tau=1$.

\begin{figure}
\rotatebox{0}{\resizebox{10cm}{13cm}{\includegraphics[0in,0.5in]
 [8in,10.5in]{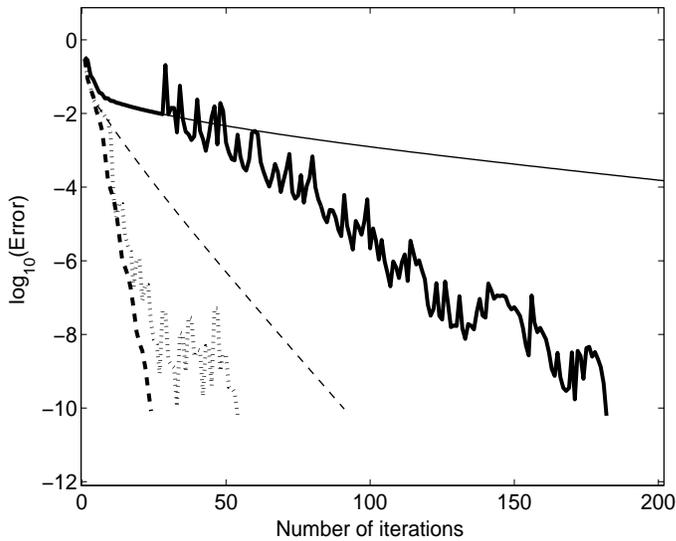}}}
\vspace{-3cm}
\caption{The evolution of the error in cases (a) and (b) of Example 4.2.
Thin solid: non-accelerated method (\ref{p2_5_04}) for
case (a); \ thick solid: method (\ref{e6_07}), (\ref{e6_08}) with $s=1$ for case (a); \
thin dashed: non-accelerated method (\ref{p2_5_04}) for
case (b); \ thick dashed: method (\ref{e6_07}), (\ref{e6_08}) with $s=1$ for case (b);
thick dotted line: Steffensen's method for case (b).
}
\label{fig10}
\end{figure}

In case $(a)$, the generalized \PM\ (\ref{p2_5_04}) takes
about 950 iterations to converge to the accuracy of $10^{-10}$. When
the mode elimination technique is applied, starting at the moment
when the error becomes less or equal to some small value (we chose
$10^{-2}$), the convergence occurs in about 180 iterations, i.e.
more than five times as fast. The evolution of the error is shown in
Fig.~\ref{fig10} by the thick solid line for the choice $s=1$; for
smaller values of $s$ up to $0.4$ which we tried, the error
evolution is similar (and the convergence is slightly faster). The
characteristic feature of this error evolution is that it is
nonmonotonic and rather irregular. This irregularity is somewhat
abated for $s<1$, in agreement with our discussion in Section 3.
Now, when we attempted to apply the Aitken's acceleration to the
generalized \PM, we observed quick divergence of the so
``accelerated" method. We actually tried various values of $n_{\rm
accel}$ and $\D \tau$ but were unable to make the iterations
converge. The reason for this is explained at the end of Section
3. In fact, by monitoring the error $\tu_n$ at every iteration, we
observed that it contains many nonlocalized modes, so that the
condition (\ref{e6_10}) of applicability of the Aitken's
acceleration is clearly violated in this case.

The corresponding results for case $(b)$ are also shown in Fig.~\ref{fig10}.
 There, the mode elimination technique
accelerates the convergence of the generalized \PM\ by about a factor of four. The error evolution is much
smoother than in case $(a)$. This appears to be correlated with the fact, which follows from our monitoring
of the error, that the latter is dominated by a single eigenmode. Consequently, condition (\ref{e6_10}) is
now satisfied, and the Steffensen's method (i.e., the generalized \PM\ with Aitken's acceleration)
also converges; see the dotted line
in Fig.~\ref{fig10}. Let us note that the irregular behavior of the error of the Steffensen's method at low
values of the error leads to a rather high sensitivity of the total number of iterations to the initial
condition. For example, we verified that if the acceleration is started when the error reaches $10^{-3}$
instead of $10^{-2}$, the Steffensen's method converges to the accuracy of $10^{-10}$ in about 30 iterations.

\begin{figure}
\vspace*{-4cm}
\rotatebox{0}{\resizebox{10cm}{13cm}{\includegraphics[0in,0.5in]
 [8in,10.5in]{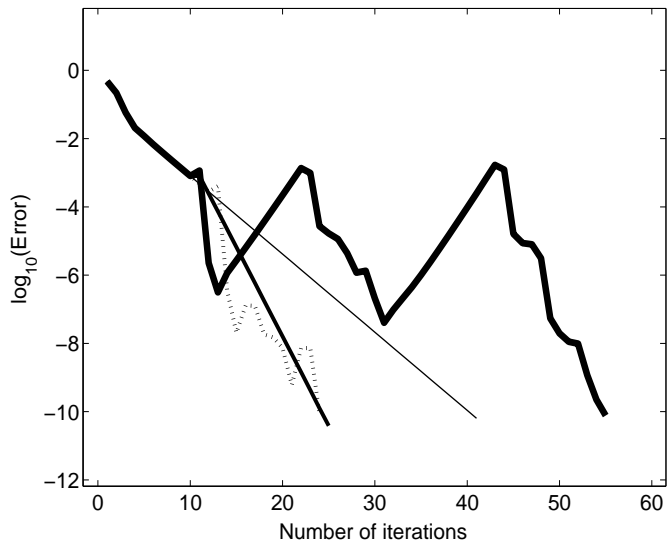}}}
\vspace{-2cm}
\caption{The evolution of the error in case (c) of Example 4.2.
Thin solid: non-accelerated method (\ref{p2_5_04}); \ thick solid: method (\ref{e6_07}), (\ref{e6_08})
with $s=1$; \ medium solid: method (\ref{e6_07}), (\ref{e6_08})
with $s=0.7$; \ thick dotted line: Steffensen's method.
}
\label{fig11}
\end{figure}

The error evolutions for case $(c)$ are shown in Fig.~\ref{fig11}.
The convergence acceleration in this case (as, actually, also in case $(b)$) is not of practical importance
because the convergence of the non-accelerated generalized
\PM\ (\ref{p2_5_04}) is quite fast (see the thin solid line in Fig.~\ref{fig11}).
Therefore, below we discuss the results for this method accelerated by the mode elimination technique
for the sole purpose of highlighting this technique's dependence on the parameter $s$.
The error evolution of method (\ref{e6_07}), (\ref{e6_08}) with $s=1$,
where the acceleration is started when the error becomes less or
equal to $10^{-3}$, is very irregular (see the thick solid line in Fig.~\ref{fig11}), and as a result, the
accelerated method takes more iterations to converge than the non-accelerated one. Moreover, the evolution
of the error also strongly depends on the initial condition and on when the acceleration is started.
For example, when we began the acceleration at the moment of the error reaching $10^{-2}$ or $10^{-4}$,
rather than $10^{-3}$, the convergence occurred in about 190 or 100 iterations, respectively. In both
cases, the error evolution curves were irregular, with several ``ups and downs". However, when we used
values $0.4 < s < 0.8$ instead of $s=1$, the behavior of the accelerated iterations greatly improved.
The optimal case of $s=0.7$ is shown in Fig.~\ref{fig11} by the medium solid line. Both the sensitivity to
the ``starting moment" of the acceleration and the irregularity of the error evolution are suppressed
for $s<1$, in agreement with the discussion in Section 3. We also applied the Steffensen's method to
this case and found it to converge in about the same number of iterations as the mode elimination method
with the optinal $s$; see the dotted line in Fig.~\ref{fig11}.

\smallskip

{\bf Example 4.3} \ In this and the following Examples, we show that
the mode elimination technique can be used to accelerate
convergence of other iterative methods. In this Example, we apply
this technique to the squared-operator method (\ref{e6_15}), which
can converge \cite{SO} to any given nonfundamental solitary wave of
the underlying stationary wave equation. It should be noted that in
\cite{SO}, the efficiency of the so accelerated squared-operator
methods (referred to there as modified squared-operator methods) was
amply demonstrated for a number of single and coupled stationary
wave equations, both Hamiltonian and dissipative. In all simulations
reported in \cite{SO}, the value of the parameter $s$ in
(\ref{e6_08}) was taken to equal 1. Therefore, below we will focus
on the dependence of the error evolution on the parameter $s$.

We apply the squared-operator methods with and without mode
elimination to finding the second nonfundamental solution of Eq.
(\ref{e6_12}). This solution for $\mu=3$ is shown in
Fig.~\ref{fig12}. In all cases considered below, we used the initial
condition \ $u_0=(1-2r^2)\,e^{-r^2}$, $r^2=x^2+y^2$ and the step
size $\D \tau=0.3$ (nearly optimal). 
As the method without mode elimination, we used (\ref{e6_15}).
The method with mode elimination is then a straightforward
modification of methods (\ref{e6_07}), (\ref{e6_08}) and
(\ref{e6_15}):
\be
\hspace*{-10cm} u_{n+1}-u_n=
\label{e6_17}
\ee
$$
- \left[ (N^{-1}LN^{-1}L_0 u)_n - \Gamma_n
  \frac{ \langle u_n, \, (LN^{-1}L_0 u)_n \rangle }{ \langle u_n, \, Nu_n \rangle }\, u_n -
  \Gamma_{{\rm slow}, \,n}
  \frac{ \langle \phi_{{\rm slow},\,n} , \, (LN^{-1}L_0 u)_n \rangle }
  { \langle \phi_{{\rm slow},\,n} , \, N \phi_{{\rm slow},\,n}  \rangle }\, \phi_{{\rm slow},\,n}
  \,\right] \D \tau,
$$
where, similarly to (\ref{e6_08}):
\be
\phi_{{\rm slow},\,n} = u_n-u_{n-1}, \qquad
\Gamma_{{\rm slow},\,n} = 1 - \frac{s}{{\cal A}_{{\rm slow},\,n} \D \tau},  \qquad
{\cal A}_{{\rm slow},\,n} = 
\frac{ \langle \phi_{{\rm slow},\,n} , \, LN^{-1}L \phi_{{\rm slow},\,n} \rangle }
  {\langle \phi_{{\rm slow},\,n} , \, N \phi_{{\rm slow},\,n} \rangle } \,.
\label{e6_18}
\ee
In both cases, with and without mode elimination, we found empirically that the
methods with $\Gamma_n$ given by (\ref{e6_16}) require the initial
condition to be closer to the exact solution than do 
the corresponding methods with $\Gamma_n=0$. On the other hand, the former
methods were significantly faster than those with $\Gamma_n=0$. Therefore, 
we initially used methods (\ref{e6_15}) or (\ref{e6_17}) with
$\Gamma_n=0$, and when the error reached a small value (we chose
$5\cdot 10^{-3}$), switched $\Gamma_n$ to the expression
(\ref{e6_16}). 
The corresponding error evolutions for the accelerated method (\ref{e6_17}) with
$s=1$ and $s=0.7$ (optimal) are shown by the thick and medium lines,
while for the non-accelerated method (\ref{e6_15}) without mode elimination, 
the error evolution is shown by the thin line. Note that the
behavior of the accelerated method with $s<1$ compared to that
behavior with $s=1$ follows the same trends as observed in Example
4.2. Namely, the error evolution for the schemes with $s<1$ is
smoother and much less sensitive to the moment when the
acceleration starts. Overall, the mode elimination is found to accelerate
the convergence by a factor between three and four, depending on the 
choice of the parameter $s$. Finally, we note that
the Steffensen's method in this case does not converge.

\smallskip

{\bf Example 4.4} \ In this last Example, we show that the
convergence acceleration technique based on mode elimination can
also be applied to the ITEM. Here we chose to present the results
for the version of this method (\ref{add1_5_01})--(\ref{e5_09})
with power normalization, but the
technique can be used as well for the ITEM with amplitude
normalization \cite{YangL06}.

For the stationary wave equation (\ref{e5_01}) written in an equivalent form:
\be
L_0 u \equiv L_{00}u - \mu u =0,
\label{add1_6_01}
\ee
the ITEM (\ref{add1_5_01})--(\ref{e5_09}) with mode elimination can be written as follows:
\be
u_{n+1}=\left[\frac{P}{\langle \hat{u}_{n+1}, \hat{u}_{n+1}
\rangle}\right]^{\frac{1}{2}} \hat{u}_{n+1},
\label{add1_6_02}
\ee
\be
\hat{u}_{n+1}-u_n= \left[ K^{-1} (L_0u)_n  -
 \gamma_{{\rm slow},\,n}\,\frac{ \langle \phi_{{\rm slow},\,n} , \, (L_0u)_n \rangle }
 {\langle \phi_{{\rm slow},\,n} , \, K \phi_{{\rm slow},\,n}  \rangle }\, \phi_{{\rm slow},\,n}
 \right] \D \tau\,.
\label{add1_6_03}
\ee
Here $K$ is a positive definite self-adjoint operator with constant coefficients 
(as, e.g., in (\ref{p2_5_06})),
\be
(L_0u)_n=L_{00}u_n-\mu_n u_n,  \hspace{0.6cm}
\mu_n=\frac{\langle L_{00}u_n, K^{-1}u_n\rangle}{\langle u_n, K^{-1}u_n\rangle},
\label{add1_6_04}
\ee
and
\be
 \phi_{{\rm slow},\,n} = u_n-u_{n-1}, \qquad \gamma_{{\rm
slow},\,n} = 1+ \frac{s}{\alpha_{{\rm slow},\,n} \D \tau},  \qquad
\alpha_{{\rm slow},\,n} = \frac{ \langle \phi_{{\rm slow},\,n} , \,
L \phi_{{\rm slow},\,n} \rangle }
  {\langle \phi_{{\rm slow},\,n} , \, K \phi_{{\rm slow},\,n} \rangle } \,.
\label{add1_6_05}
\ee
We apply the methods without and with mode elimination --- (\ref{add1_5_01})--(\ref{e5_09}) and
(\ref{add1_6_02})--(\ref{add1_6_05}), respectively, ---
to Eq. (\ref{p2_6_02}) with $V_0=4$ and $P=1.94$, whose solution looks similar to the
top solution in Fig.~3 of \cite{gP}.
The corresponding propagation constant
$\mu=5.01$  is close to the bandgap edge, and the ITEM without mode elimination converges
slowly; see the thin line in Fig.~\ref{fig14}. 
In all simulations, we took $\D\tau =1$ and the operator $K$ of the
form (\ref{p2_5_06}) with $c=1$, which yielded the (nearly) optimal convergence rate of the ITEM
(\ref{add1_5_01})--(\ref{e5_09}). The error evolutions for the ITEM
(\ref{add1_6_02})--(\ref{add1_6_05}) with mode
elimination are shown in Fig.~\ref{fig14} by the thick and medium lines.
As in Examples 4.2 and 4.3, the scheme with mode elimination
provides a severalfold improvement to the convergence rate of
the ITEM. Also as in those Examples, the error evolution with
$s<1$ is more regular than that with $s=1$.

\bigskip

Thus, from the last three examples, we conclude that in those cases
when the iterations converge slowly and their acceleration is highly
desirable, the mode elimination method provides a
considerable improvement of the convergence rate (by a factor of
several times). Taking $s<1$, so that only part of the mode
$(u_n-u_{n-1})$ would be eliminated, usually results in smoother
convergence; however, the choice $s=1$ still yields a considerable
improvement of the convergence rate in comparison with that of the
non-accelerated iteration  method. For these slowly
convergent cases, the Steffensen's method, based on the Aitken's
acceleration, often diverges.

\smallskip

{\bf Remark} \ In those cases when the step size $\D \tau$ is nearly optimal, the error is expected to
be dominated by two eigenmodes, corresponding to $\Lambda_{\max}$ and $\Lambda_{\min}$, since
\be
(1+\Lambda_{\max}\D \tau) \,\approx\,
-(1+\Lambda_{\min}\D \tau)
\label{e6_19}
\ee
for this $\D \tau$ (see (\ref{e5_03}) and (\ref{e5_05})).
Then it seems logical that one would need to eliminate {\em both} of these eigenmodes, 
which are proportional
to $(u_n-u_{n-2})$ and $(u_n-2u_{n-1}+u_{n-2})$, respectively. We found, however, that although this does
result in a smoother error evolution than the elimination of just the single mode $(u_n-u_{n-1})$, it
does not yield any consistent improvement of the convergence rate compared to the latter case.


\section{Summary}

In this work, we obtained the following results.

In Section 2, we compared the linearized operators of the
generalized \PM\ and the ITEM with power normalization. In particular, we showed that while the 
``primordial" part of those operators has the same form (\ref{e5_11}), 
their nonlocal parts involving the inner products are different, leading to the eigenvalues
of the corresponding operators being different. In our simulations we observed that the generalized 
\PM\ converges to fundamental solitary waves faster than does the ITEM (although we
could prove this rigorously only for equations with power-law nonlinearity). On the other hand,
in those (rare) cases when both methods converge to a nonfundamental solitary wave, we produced an
explicit example where the ITEM is faster.

In Section 3, we proposed a new technique, which we referred to as the mode elimination.
One application of this technique is that it can obtain nonfundamental solitary
waves, for which the generalized Petviashvili method would otherwise diverge. 
The corresponding iteration scheme is given by Eqs. (\ref{e6_03}) and (\ref{e6_04}).
In Example 4.1 in Section 4, we demonstrated that
this technique can be superior to an alternative,  squared-operator, technique \cite{SO} when
applied to finding {\em lowest-order} nonfundamental solutions.
However, for finding higher-order solutions, the technique of Ref.
\cite{SO} appears to be more practical. 

As a {\em more important application} for the mode elimination technique, 
we showed that it can accelerate the convergence
of various iteration methods. This acceleration is most significant (by a
factor of several times) in those cases when it is most needed,
i.e., when the convergence of the non-accelerated method is slow. 
The iteration schemes implementing this technique are: Eqs. (\ref{e6_07}), (\ref{e6_08})
for the generalized \PM; Eqs. (\ref{e6_17}), (\ref{e6_18}) for a squared-operator
method (see also Ref.~\cite{SO}); and Eqs. (\ref{add1_6_02})--(\ref{add1_6_05}) for the
ITEM with power normalization.

\section*{Acknowledgement}
The work of T.I.L was supported in part by the National Science Foundation under
grant DMS-0507429, and the work of J.Y. was supported in part by The Air Force Office of
Scientific Research under grant USAF 9550-05-1-0379.


\newpage

\vspace{-2cm}

\vspace*{-2cm}

\vspace*{-3cm}

\vskip -3 cm

\vskip 3 cm

\newpage

\begin{figure}
\rotatebox{0}{\resizebox{10cm}{13cm}{\includegraphics[0in,0.5in]
 [8in,10.5in]{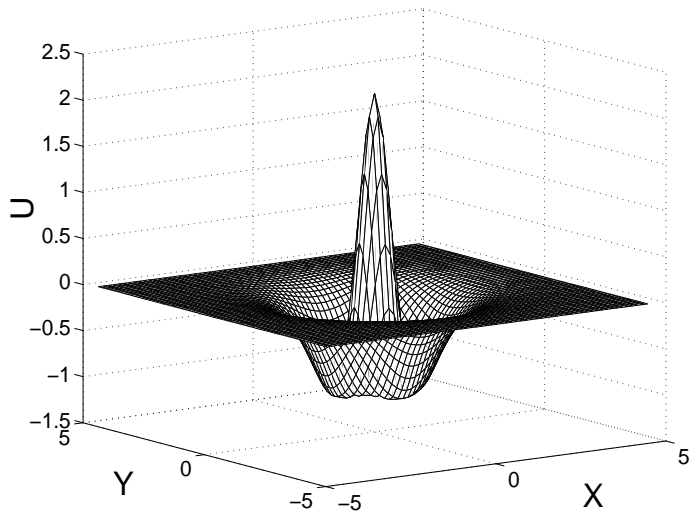}}}
\vspace{-2cm} \caption{The second nonfundamental solution of Eq.
(\ref{e6_12}) with $\mu=3$. }
\label{fig12}
\end{figure}

\vskip 3 cm

\begin{figure}
\vspace*{-4cm}
\rotatebox{0}{\resizebox{10cm}{13cm}{\includegraphics[0in,0.5in]
 [8in,10.5in]{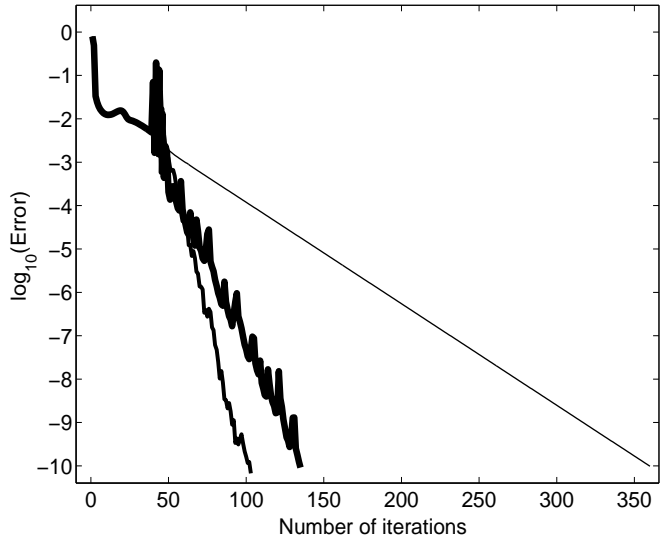}}}
\vspace{-2cm}
\caption{The evolution of the error in Example 4.3; in all cases, the application of
(\ref{e6_16}) and/or acceleration is begun when the error norm reaches $5\cdot 10^{-3}$.
Thin line: method (\ref{e6_15}) (no mode elimination); \ thick line: method (\ref{e6_17})
with $s=1$; \ medium line: method (\ref{e6_17}) with $s=0.7$.
}
\label{fig13}
\end{figure}

\newpage

\begin{figure}
\vspace*{-4cm}
\rotatebox{0}{\resizebox{10cm}{13cm}{\includegraphics[0in,0.5in]
 [8in,10.5in]{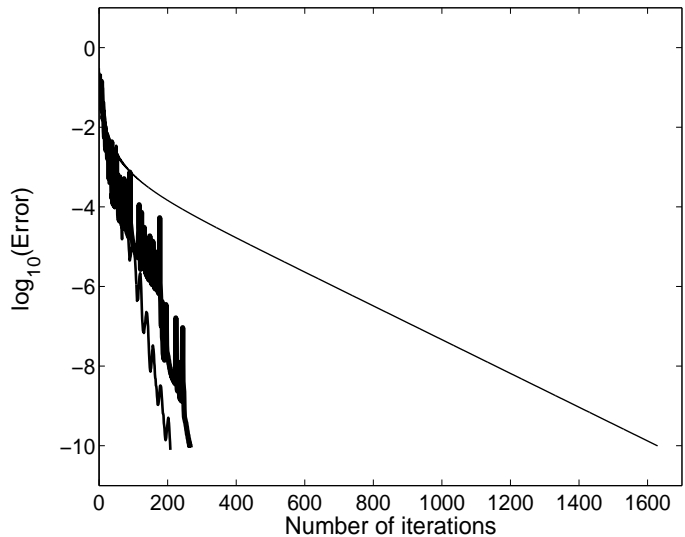}}}
\vspace{-2cm}
\caption{The evolution of the error in Example 4.4.
Thin line:  optimally accelerated (with respect to parameter $c$ in operator $K$, see text) ITEM
(\ref{add1_5_01})--(\ref{e5_09}) without mode elimination. Thick and medium lines: ITEM
(\ref{add1_6_02})--(\ref{add1_6_05}) with mode elimination with $s=1$ (thick) and $s=0.7$
(medium). The application of mode elimination begins at the first iteration.
 }
\label{fig14}
\end{figure}

\end{document}